# Investigating Political Participation and Social Information Using Big Data and a Natural Experiment

**Scott A. Hale[1], Peter John[2], Helen Margetts[1] and Taha Yasseri[1]***
[1] **Oxford Internet Institute, University of Oxford, Oxford, UK**
[2] **School of Public Policy, University College London, London, UK**

**Abstract** Social information is particularly prominent in digital settings where the design of platforms can more easily give real-time information about the behaviour of peers and reference groups and thereby stimulate political activity. Changes to these platforms can generate natural experiments allowing an assessment of the impact of changes in social information and design on participation. This paper investigates the impact of the introduction of trending information on the homepage of the UK government petitions platform. Using interrupted time series and a regression discontinuity design, we find that the introduction of the trending feature had no statistically significant effect on the overall number of signatures per day, but that the distribution of signatures across petitions changes—the most popular petitions gain even more signatures at the expense of those with less signatories. We find significant differences between petitions trending at different ranks, even after controlling for each petition's individual growth prior to trending. The findings suggest a non-negligible group of individuals visit the homepage of the site looking for petitions to sign and therefore see the list of trending petitions, and a significant proportion of this group responds to the social information that it provides. These findings contribute to our understanding of how social information, and the form in which it is presented, affects individual political behaviour in digital settings.

*\* Authors are in alphabetical order and made equal contributions.*

# Introduction

Social information—information about what others are doing or have done—has always been a powerful influence on political behaviour. But in digital settings people are exposed to more social information than ever before. When confronted with an item of news or information, or invited to participate in a political mobilization or campaign, users of social media know in real-time how many other people have 'liked', shared, downloaded, 'favourited', 'retweeted', reviewed or commented upon that item. They may be able to read other individuals' comments and feedback, and they may be recommended another item on the basis of the behaviour of other people. Indeed real-time social information is so ubiquitous that it is no longer easy to distinguish its effect on individual behaviour, given that it is the default setting for most activities online. However, the more that political participation takes place online—and therefore under conditions of greater social information—the more important it becomes to understand the influence of social information at the micro level.

Platform changes can provide a valuable opportunity to isolate and understand social information effects on contemporary political behaviour, and thereby provide valuable insight into the dynamics of contemporary political mobilization. Platform changes can provide conditions for a natural experiment as visitors to the site just before and after the change experience the change as if it were random (Dunning, 2012). Under these conditions, there is no reason to expect different behaviour before and after the change that cannot be directly attributed to the change. This paper analyses the effect of one such change—the introduction of trending information onto the homepage of the UK government petitions platform—and its impact on petition signing. It uses data pertaining to internet-based mobilizations around petitions, generated from the current electronic petitions platform in the UK, which was developed by the UK Cabinet Office for the incoming Coalition Government in 2010 and launched in August 2011. The data collected on the site since its launch provides an example of real-time transactional data on this popular form of political participation for a whole population (N = all), and thereby represents so-called 'big data'. First, we review some recent literature on social information and participation and relate it to online settings. Second, we provide some background on petitioning as a form of collective action and outline the distinctive features of the e-petition platform we examine here. Third, we present the petitions data and analyse it with interrupted time series analysis and a regression discontinuity design. We discuss the key findings from the data, which suggest that a small but significant group of users (who we label 'aimless petitioners') are indeed influenced by the trending information. Fourth, we use site analytics data from the petitions platform to understand further the size, constituency and behaviour of this group of users. To conclude, we discuss the implications of our findings for political science research into political participation in digital settings and for those who design portals aimed at citizen engagement.

# Social information and Digital Media

The term social information originates from social psychology, where social information processing refers to the study of the informational and social environments within which individual behaviour occurs and to which it adapts (Pfeffer and Salancik, 1978; Salancik and Pfeffer, 1977). Social information helps people decide what they are going to do with reference to a wider social group, which has the potential to activate people's social norms. Potential participants construe this information as representing the behaviour of a 'generalised other', or a social aggregate (Schelling, 1978), and take it into account when they are deciding whether and how to participate. Social information has been shown to affect charitable giving and willingness to participate in public goods provision (see Andreoni, 2006; Cialdini and Goldstein, 2004; Goldstein et al, 2008; Frey and Meier, 2004; Shang and Croson, 2009). Economists have studied the effects of social information on people's willingness to undertake pro-social behaviours, in particular in making charitable donations, where cooperation that is dependent on social information giving evidence of the contributions of others has been labelled conditional cooperation. Social information may crowd out contributions or crowd them in, with evidence pointing to the latter because of conformity, social norms or reciprocity (see Frey and Meier, 2004, 1721). People are more likely to contribute to a campaign if they are provided with information that other people are doing so, and this effect increases with larger numbers of additional participants (Frey and Meier, 2004; Fischbacher and Gachter, 2010). It has also been shown that people are likely to increase their contribution (by donating more money, for example) if they know that other people are increasing the size of their commitments (Croson and Shang, 2008; Shang and Croson, 2009). In sociology and political science, experimental studies have shown the importance of social information on people's willingness to contribute to public goods by undertaking activities such as recycling (Goldstein et al., 2008; Schultz, 1999; Cotterill et al., 2010) and voting (Gerber, Green and Larimer, 2008). Social information provides a crucial signal of viability for a mobilization, evidence of whether or not it has reached or will reach 'critical mass' (Marwell and Oliver, 1993) and hence the potential benefits of joining, thereby altering the incentives of individuals to participate. Research has identified one million signatures as a possible tipping point for online petition-based mobilizations: if potential participants know that more than one million people have already participated, they are more likely to participate themselves (Margetts et al., 2011).

When social information is provided precisely and in real-time, as facilitated by online social media, we expect that these effects on mobilization will be all the more profound and intractable. For example, we might expect to see in a political context the same kind of effects observed in cultural markets, where Duncan Watts and colleagues (Salganik et al. 2006a and 2006b; Salganik and Watts 2009) carried out a series of experiments to show how changes to real-time information feedback about cultural artefacts (songs) changes the way that people view the artefacts' quality, leading popular songs to become more popular and unpopular ones less so. Furthermore, internet and mobile technologies also reduce the transaction costs of political participation: they facilitate 'micro-donations' of time, effort and money towards political causes, ease the making of small monetary donations to political causes (texting a keyword to donate £3 to a disaster relief effort, for example), and extend the lower end of the 'ladder' of participation (Margetts et al., forthcoming). As the costs of participation plummet, the cost-benefit analysis facing someone deciding whether to participate changes (Lupia and Sin, 2003), and social information is likely to be proportionately more important an influence when deciding whether to make such a micro-donation, as other influences diminish. Indeed recent research investigating the relationship between internet use and participation has shown that where costs of participation are very low, interest in politics reduces in importance as a causal factor of participation and that skilled internet users do not need to be motivated or interested in politics in order to participate online (Borge and Cardenal, 2012).

In the next section, we outline the changing form of political participation in online settings, in particular the rise of petitioning, which forms the context of our study.

**Petitions Platforms**

Signing petitions has long been one the more popular political activities, leading the field for participatory acts outside voting, and with other social benefits such as civic mindedness (Whyte et al., 2005) ascribed to it in addition to its potential to bring about policy change. In the UK the right to petition the king goes back to medieval times and Richard II. Petitions were widely used by the 18[th] century and were a key mechanism in the campaign for parliamentary reform in the early 19[th] century (Fox, 2012), when petitioning was a popular activity in the US (Carpenter 2003). After a period of decline through the 20[th] century, petitioning received a renewal of popularity in the 21[st], with the availability of electronic petitions that could be created, signed and disseminated on the internet and, more recently, social media. Although there are start-up costs in getting to know the platform for people who initiate petitions, organizers can find supporters more easily, rather than having to canvass them door-to-door or approach people in the street. For those wishing to sign petitions, the search costs are far lower: they may sign a petition instantly on receipt of an email or post on a social networking site, or go to one of the large number of petition platforms and look for a petition to sign, rather than having to wait until they encounter a petition in the course of their other online activities. Furthermore, ever petition signer is now potentially a petition organizer given the ease with which petitions can be disseminated to one's contacts via social media. With these reduced transactional costs and the ease of coordination, we would expect e-petitions to be highly successful in the age of social media, and indeed they have become so, and are one of a growing portfolio of internet-based democratic innovations (Smith 2009). Both governments and NGOs, such as Avaaz and 38 Degrees, have made widespread use of e-petitions and accordingly have received accolades for their democratic contribution by academic commentators (Escher 2011; Chadwick 2012). A growing number of governments have implemented petition platforms, notably the UK, the US and Germany. While the German e-petition platforms have received extensive analysis by political scientists (see Lindner and Riehm 2011; Jungherr and Jurgens 2012), the UK petition platforms have received rather less attention in recent political science research, with the exception of Wright (2012) and Fox (2012) and our own work (Margetts et al., 2011, Hale et al., 2013, Yasseri et al., 2014).

The first e-petitions platform in the UK was launched on the No.10 Downing Street website in November 2006. Created by the social enterprise MySociety, the platform received more than 8 million signatures from over 5 million unique email addresses over the course of its lifetime from November 2006 until 4 April 2011.[1] The site allowed any user with a valid email address to create a new petition or sign an existing one. For the No. 10 site, prospective petitioners were told that if their petition achieved 500 signatures, they would receive an official response from government. There were no other official measures of success, although one petition on road pricing did succeed in raising 1.8 million signatures in 2007 and was widely regarded as influential in getting the government's policy reversed. With regard to social information, this first site provided the total number of signatures each petition had and the names of the first 500 people to sign each petition. This first site was closed by the incoming Conservative-Liberal Democrat Coalition Government in March 2011. The Cabinet Office launched a new petitions site in August that year, initially on the direct.gov portal (which eventually became the new www.gov.uk portal in the autumn). The new petitions site, which we study in this paper, also shows the total number of signatures per petition, but does not show the names of petition signers except for the name of the petition creator. The site also provides new measures of success for petitions: the bar for an official response from government was initially unclear from the site, although the majority of petitions with over 10,000 signatures do receive a response with the prefix 'As this e-petition has received more than 10,000 signatures, the relevant Government department have provided the following response', and this assurance was eventually formalised. More importantly, in the early days of the Coalition, Prime Minister David Cameron promised that petitions obtaining more than 100,000 signatures would generate a parliamentary debate on the issue and a handful of petitions have passed this bar.

**Changing Social Information: The Introduction of Trending Petitions**

Both the UK government petition platforms displayed social information, in terms of the number of other people who had signed any given petition. It was also possible to sort the list of petitions by popularity that is, by the number of signatures, on both the old and the new sites, and indeed popularity is the default sorting order on the new site. In 2012, social information was emphasised when a change was made to the site to show trending petitions on the homepage: a list of the six most popular petitions in terms of total number of signatures within the last hour. This change means that these petitions are flagged for attention to any visitor to the homepage, with social information automatically provided—regardless of whether the visitor seeks it out—and acting as a potential driver for signature growth. Therefore, this change to the platform provides the opportunity to test the hypothesis that social information of this kind (that is, an indicator of popularity and rate of change, given that the number of signatures in an hour indicates movement on the petition) acts to encourage participation for those petitions for which it is provided. We would expect that those petitions for which trending information is displayed to receive disproportionately more petitions than those that do not.

---

[1] http://www.mysociety.org/projects/no10-petitions-website/

Such expectations about feedback depend on expectations about the shape of social responses in collective action on the internet, which depends on how people get to know what other people are doing politically. From the theoretical discussion reviewed earlier, we consider two alternative hypotheses:

H1: Trending information will lead to more overall signatures on the petitions platform

H2: Overall signatures on the petitions platform will remain constant, but trending petitions will receive more signatures at the expense of non-trending petitions

We are able to test these hypotheses because, as indicated above, we have collected all the signatory data for the platform before and after the change. This means that the platform change creates a natural experiment, allowing us to examine behaviour before and after the change in a regression discontinuity design.

**Data**

Online petitions are created, disseminated, circulated, and presented online. They are regularly disseminated online via social media platforms, particularly Facebook and Twitter. Although policy-makers may discuss responses in offline contexts, such responses are disseminated online. So both successful and unsuccessful mobilizations that form around petitions leave a complete digital audit trail of the real (rather than reported) actions of the entire population of signers of each petition, however large or small. In contrast, paper-based petitions left a rather incomplete record given that unsuccessful petitions, which were not presented to the government or legislature, were unlikely to be recorded, although successful ones can now be transcribed to digital form and analysed (see Carpenter, 2003 for such an analysis of anti-slavery petitions in the US). The digital imprint of electronic petitions can be harvested to provide so-called 'big data', real-time transactional data for a complete population which can be analysed to provide valuable understanding of the dynamics of contemporary participation. At the same time, such data provide a number of challenges to analysis, given that we have no demographics or survey responses with which to cross-tabulate, nor any indication of the past or future activities of the users of the petitions site.

We earlier accessed the UK Government's first petition website (petitions.number10.gov.uk) daily from 2 February 2009 until 4 April 2011, when the site closed and no further signatures could be added, with an automated script. Each day, the number of overall signatures to date on each active petition was recorded. In addition, we collected the name, the text, launch date, and the category of the petition. Overall, 8,326 unique petitions were tracked, representing all publicly available petitions active at any point during the collection period. Initial analysis of this data after the site closed revealed the importance of the first day in the future success of a petition (Hale, et al., 2013) and suggested that more frequent scraping of the data could deliver a finer-grained analysis.

For this reason, when the new petitions site was launched in August 2011, we set a new automatic script to scrape it every hour, recording the same details as for the previous site. Our second dataset currently contains hourly data points for all the petitions (19,789) submitted to the new site between 5 August 2011 and 22 February 2013. This data has made it possible to examine the different patterns of growth in the 20,000 mobilization curves that we have data for and identify the distinctive characteristic of those mobilizations that succeed and those that fail (Yasseri et al., 2014; Margetts et al., 2015). Such an analysis, using data that has rarely before been available to political science researchers, can tell us a great deal about petitioning and indeed, about the nature of collective action itself in a digital world.

Our previous work has analysed the data from both UK petition platforms to show that most petitions fail (Hale et al., 2013; Yasseri et al., 2014; Margetts et al., forthcoming). Consistently across the two datasets, only 5 per cent of petitions obtained 500 signatures (sufficient to gain an official response on the first platform). On the second platform, 4 per cent received 1,000 signatures or more, 0.7 per cent attained the 10,000 signatures required to receive an official response, and 0.1 per cent attained the 100,000 signatures required for a parliamentary debate. Once again, we found the first day was crucial to achieving any kind of success. Any petition receiving 100,000 signatures after three months needed to have obtained 3,000 within the first 10 hours on average. To quantify further the fast decay in the popularity of the petitions, we modelled the number of signatures over time using a modulated multiplicative process (Yasseri, et al., 2014). We observe that the modulating parameter, which we label the 'outreach factor', decays very rapidly, reducing to 0.1% within 10 hours of the launch of a petition.

**Method**

As noted above, in March 2012, the UK Cabinet Office introduced a change to the e-petitions site that altered the information environment of prospective petitioners, by introducing a 'Trending e-petitions' facility on the homepage, providing potential signatories with a new kind of social information about which petitions were successful and how many other people had signed. We assume that such a change is exogenous to political participation itself; so, the time directly before and after the change is as if random. The fact that we captured data from this site both before and after this change provides us with a natural experiment (avoiding 'bundling' problems, Dunning, 2012: 302) whereby we can test the effect of this change using a variety of methods, including interrupted series and regression discontinuity design.

First, we tested the hypothesis (H1) that the trending facility would increase the overall number of signatures on the site, by attracting visitors to the most popular petitions. We calculated the daily number of signatures to all petitions for a period of 22 months, including 10 months before the change and 12 months after it (Figure 1). We calculated an autoregressive moving-average time series under the null hypothesis that the introduction of trending petitions had no effect on the daily number of signatures, with the smoothing time windows of width $w$ =7, 30, and 90 days. We found that indeed the change had no significant effect, and does not reject the null hypothesis (Figure 2). We can say, therefore, that the introduction of the trending facility on the website had no significant effect at the aggregated level in terms of the number of signatures per day on the site, contrary to our first hypothesis (H1) that trending information would lead to more overall signatures on the platform.

**Figure 1.** Aggregate number of daily signatures on the UK Cabinet Office petitions platform before and after introduction of the trending petitions facility. Dashed line (March 2012) shows the point at which the trending petitions facility was introduced. Data has been smoothed within time windows with the width w.

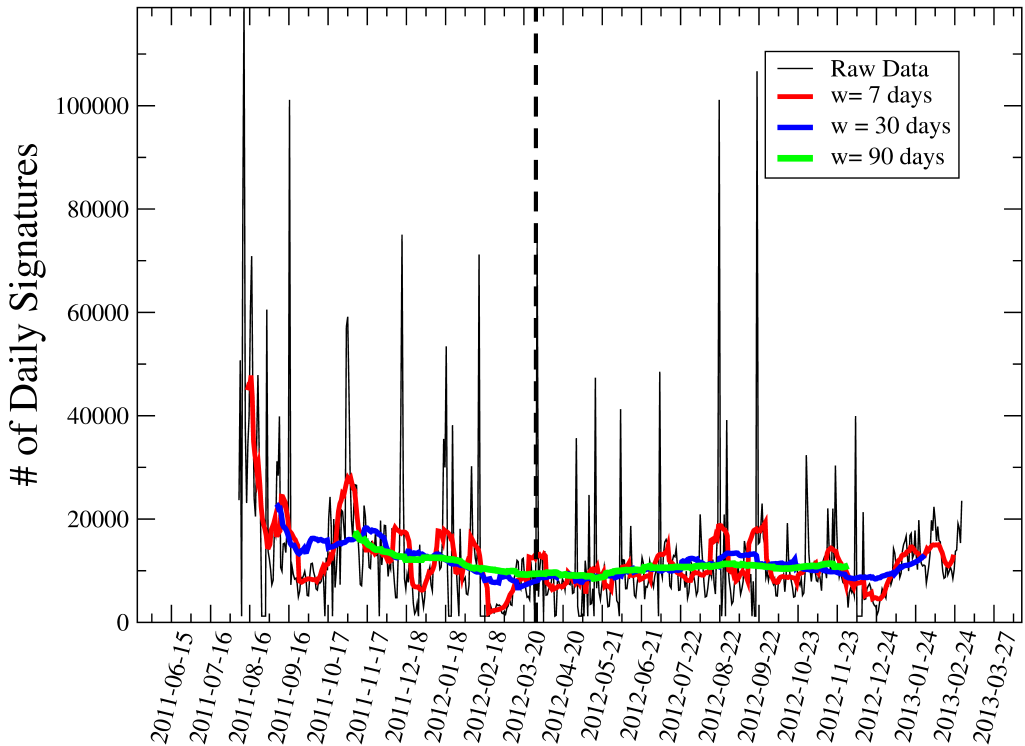

If the introduction of the trending petitions facility did not affect the overall number of signatures on the site, did it change the distribution of signatures across petitions, validating hypothesis H2? Do the most popular petitions that appear in the trending box become more popular than they would have been had the facility not been introduced? We attempted to answer these questions by calculating the Gini coefficient (the measure of inequality of a distribution) for all signatures on all active petitions before and after the change, to see whether the distribution had become more unequal. We tested the distribution of signatures over petitions for windows of four days, calculating the Gini coefficient of the distribution within the time window. We then averaged for a period of three months before the change and three months after. The result of this analysis is shown in the right panel of Figure 2. It shows clearly that after the trending petitions facility was introduced, the Gini coefficient increased significantly. We can conclude then that the introduction of trending petitions on the homepage changed the distribution of signatures across petitions. The increase in the Gini coefficient indicates that signatures were more concentrated on a small number of petitions after the design change than they were before.

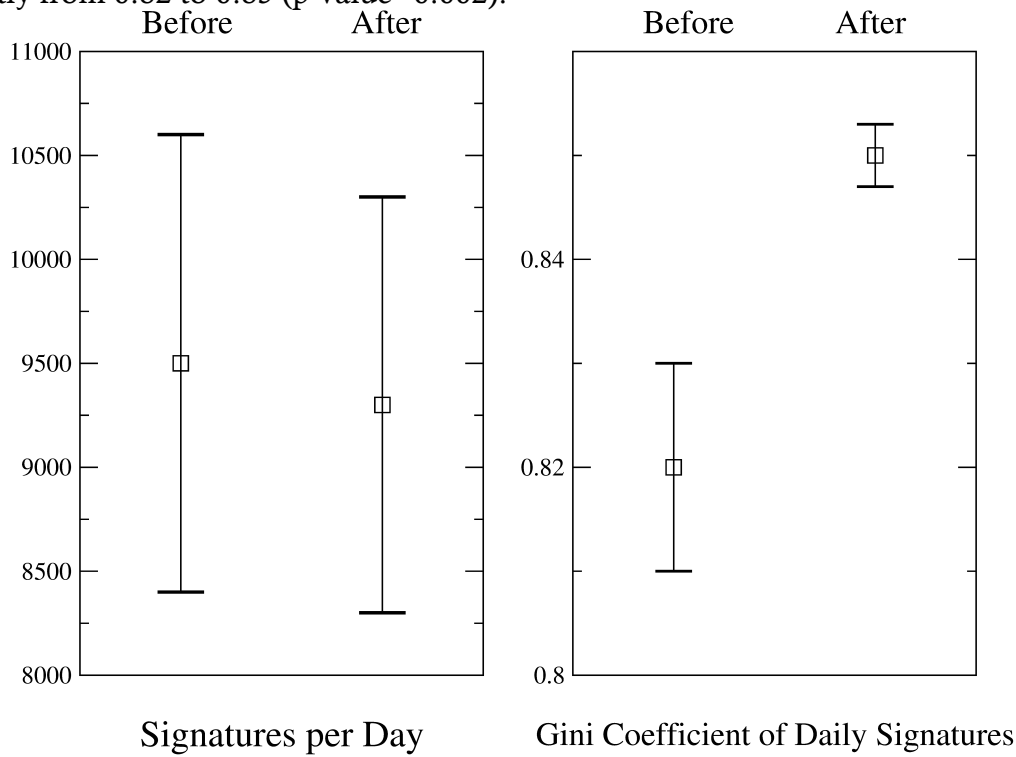

**Figure 2.** The effect of a change in design on daily signatures. Left panel: The number of daily signatures is shown; the average number over a 3-month period was 9,500 before the addition of the trending petitions facility and 9,300 after, a non-significant difference (p-value=0.2). Right panel: After the trending petitions facility was introduced, the Gini coefficient increased significantly from 0.82 to 0.85 (p-value=0.002).

**Position Effect**

We now ask to what extent petitions in different positions on the homepage were affected. Here there are aspects of the presentation of social information that vary by chance, which have to do with the ordering of when people see how many other people have signed. In our case, the change in the Gini coefficient indicates that the distribution of signatures across petitions has changed. Presumably the trending petitions have captured more signatures at the expense of non-trending petitions. In this part of the analysis, this hypothesis can be tested directly with a regression discontinuity (RD) design, which is where a regression on the outcome using observations around the cut point can estimate the impact of the variable of interest (Dunning, 2012: 63).

At this point, it is worth giving more detail about the design of the petitions site. Trending petitions appear on the homepage in a two-column ranked-ordered list. The petition with the most signatures in the last hour is in the top-left of the grid (position 1), while the petition with the second most signatures in the last hour is the top-right of the grid (position 2). Position 3 is in the left column of the second row and position 4 is in the right column of the second row, etc. The top trending petitions are shown in a 2-column, 3-row grid with the option to click a link to expose petitions in positions 7-12. Each user may only sign a petition once, and users do not know how many others will sign the same petition or another petition. It is therefore not possible for one user to directly control the position at which a petition trends on the homepage (or if it trends at all). The important difference in outcomes is between petitions being placed in different positions, and we can compare the signatures petitions in adjacent positions receive while trending on the homepage.

Our analysis here differs from a classic RD design in three ways. First, the number of signatures a petition receives is a discrete variable; so, we use the generalization of the RD design for discrete values used in Narayanan et al. (2011). Second, the position of a petition is determined by the number of signatures in the hour prior to trending while we measure the number of signatures in the hour while trending, and we cannot assume complete independence between these values (something we address further below). Finally, we only observe trending petitions once an hour while the list is updated continuously in real time.

Figure 3 shows the difference in the raw number of signatures petitions in adjacent positions receive within one hour. Petitions in position one, for instance, receive 66-74 more signatures in the hour while trending than petitions trending in position two. This effect diminishes rapidly for lower positions.

**Figure 3.** Comparisons of raw signatures in adjacent positions. The graph shows 95% confidence intervals in the difference in the raw number of signatures petitions trending in adjacent positions receive while trending.

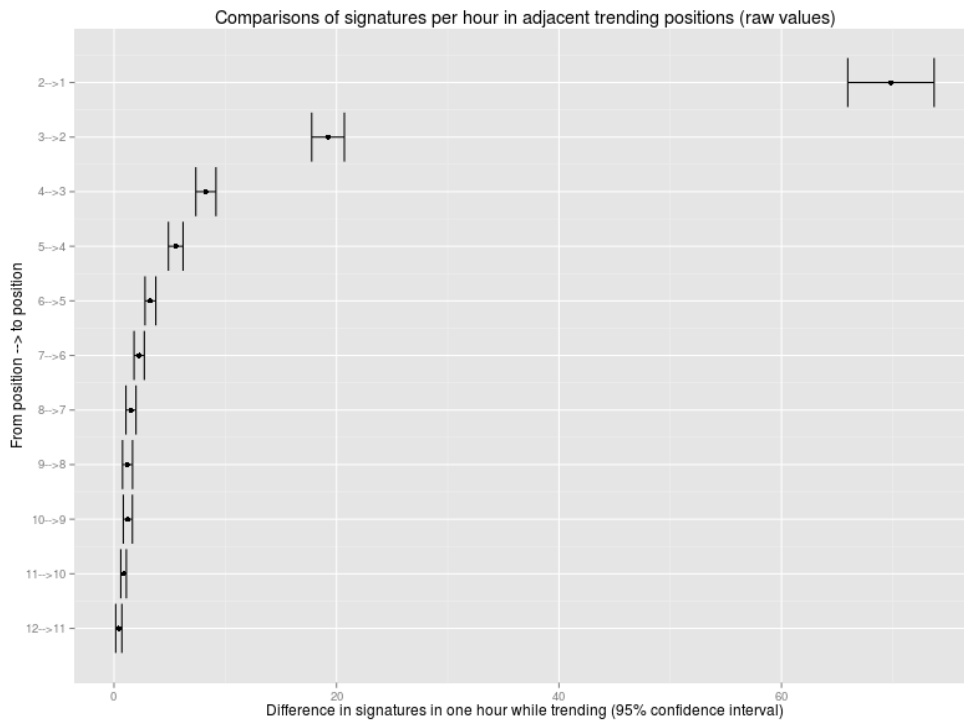

This analysis, however, does not take into account the differences among individual petitions. To control for these differences, we calculate the mean number of signatures for each petition over a number of hours before it trends. Figure 4 shows a comparison of the differences from these means for adjacent positions. The means for the figure are calculated over a window of 18 hours prior to each petition trending. The results are stable for windows of other sizes larger than this.

**Figure 4.** Comparisons of difference from 18 hour means in adjacent positions. The graph shows 95% confidence intervals in the difference between petitions trending in adjacent positions simultaneously, but first subtracts from each petition its mean number of signatures for the 18 hours prior to trending in order to control for the individual variation of petitions.

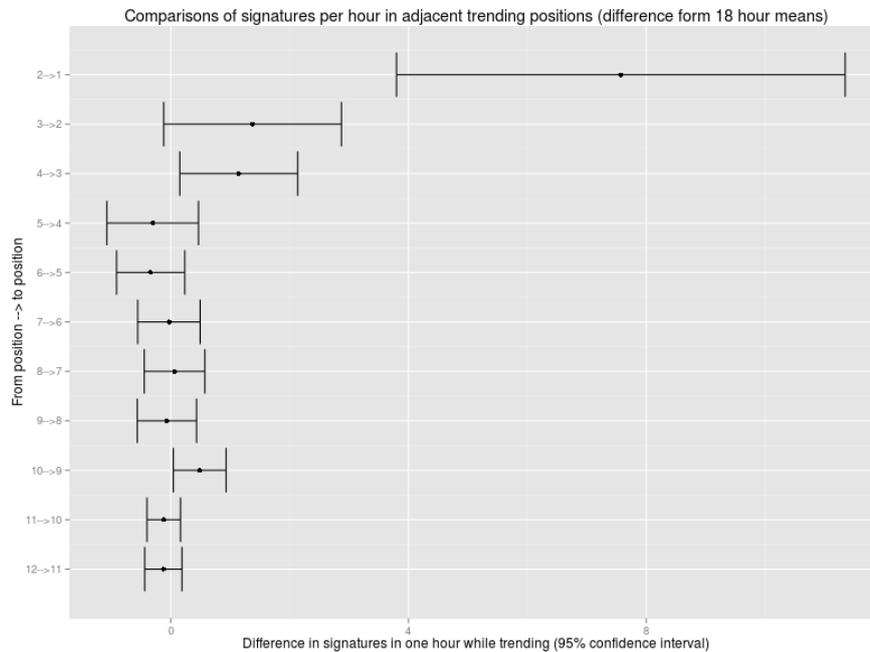

Within the limitations detailed above, our analysis shows that the trending petitions facility does concentrate attention and signatures on the first few ranked petitions – but also that it does so differentially according to the positioning of the petitions. That is, we find that the petition trending in the first position receives 3.8 to 11.3 more signatures above its per hour mean than the petition trending in the second position. Similar, a move from position 4 to position 3 is associated with 0.2 to 2.1 more signatures above the petition's eighteen-hour mean.

    The effect quickly dissipates for petitions in lower-ranked positions. So, while the trending facility does concentrate signatures, it does so only for petitions in the first few positions. Strangely, we see significant effects for moving from position 2 to position 1 and from position 4 to position 3, but not for moving from position 2 to position 1. We suspect the two-column layout is responsibility for this oddity given the findings of eye-tracking studies with web search results in the field of Human-Computer Interaction (e.g., Buscher, Cutrell, and Ringel Morris, 2009). Users of search engines usually fixate on the results in the top position and then skim down the left side of the page (at least for users with left-to-right languages). This tendency to skim down the left side of the page means petitions in the left column (positions 1, 3, and 5) may stand out more than petitions in the right column (positions 2, 4, and 6) and likely explains why a move from position 3 to position 2 has no significant effect at a 95% confidence level.

    Overall, our results provide an interesting test for the provision of social information and find that the trending petition information concentrates attention to the top few ranked petitions. We expected to find a significant difference between petitions in positions 6 and 7, because position 6 appears in the bottom right of the trending petition information while viewing position 7 requires a further click to 'see additional trending petitions.' However, attention is so concentrated on the first few positions that we find no significant difference between petitions in positions 6 and 7.

    The analysis of positions in combination with the number of overall signatures per day

remaining constant and the change in the Gini coefficient show that the addition of the trending petitions facility means that trending petitions receive more signatures and that these signatures come at the expense of signatures to other petitions on the site (thus the change in the Gini coefficient and the overall number of signatures per day not changing), validating H2. Simply put, the rich get richer, and the poor get poorer. Such a result shows evidence of the heavy-tailed distributions (sometimes referred to 'power-laws') that crop up so often in internet-based distributions, where social information is available about which online initiatives are popular. However, in this particular case, we found the implications of this result to be somewhat surprising. It suggests that some people come to the homepage of the petitions website looking for petitions to sign, or that they come to a specific petition on the site and then move on to the homepage looking for other petitions that interest them. These people have a zero-sum attention capacity: they will sign a certain number of petitions, but this number does not appear to have varied before and after the trending facility was introduced; so, if a particular petition attracts their attention, they will sign that one at the exclusion of another that they otherwise might have signed. It is worth noting that the design of the petitions site only lets users sign petitions one at a time, and users must re-enter their details (name, postcode, email address) separately for each petition they wish to sign. This evidence that people are coming to petition sites just to 'find something to sign' (rather than coming to sign a petition on a specific issue) suggests a general desire for political engagement—the 'aimless surfing' identified by Borge and Cardenal (2011)—without a firm view as to what the engagement should be about. This group are not completely aimless surfers however, because they know what they want to do, that is, to sign a petition. We might call this group 'aimless petitioners', people who want to do a 'little bit of politics' and sign a petition, but are not yet sure what about and we investigate their behaviour in more depth in the next section.

**Identifying the 'Aimless Petitioners'**

In the previous sections, we identified a significant change in the Gini coefficient of signatures per petition, but of a small magnitude (changing from 0.82 to 0.85). The largest possible change in magnitude, however, is limited by the number of visitors to the site exposed to the trending information, which is only on the homepage. In this section, we use site analytics data to understand what percentage of visitors is exposed to the trending information on the homepage. This allows us to compute that the largest possible shift in the Gini coefficient is only 0.4, and thus, the 0.3 magnitude shift we observed indicates that a large portion of users exposed to the trending information are affected by it. We term this group of users who go to the homepage and are influenced by the trending information as 'aimless petitioners' and investigate the behaviour of these users.

We cannot match the analytics data (which relates to user visits to the site) with the petitions data (which relates to petition signatories), and in any case do not have the analytics data for the time that the platform was changed (our analytics data is limited to the period of December 2012 to April 2014), but we can use it to look at the behaviour of users of the site who visit the homepage, and therefore see the trending information. First we look at the overall traffic sources to the website. Close to 40 per cent of all the traffic is directed from Facebook (two thirds of which is from mobile Facebook). The share of traffic directed from Facebook has been gradually increasing over the period from January 2013 to March 2014 (shown in Figure 5). To further understand user behaviour on the petitions website, and specifically to characterise the 'aimless petitioners', we analysed the behaviour flow of the users, in order to investigate how many users see the trending petitions information on the homepage and from where those users come.[2]

---

[2] To check the stability of the forthcoming measurements, we repeated them for a shorter period of 6 months starting from December 2012. The results are all broadly similar with differences only in the decimal digits, and we therefore conclude that the results are stable at the reported precision.

**Figure 5. Share of different sources of** referral traffic to the petitions website over time.

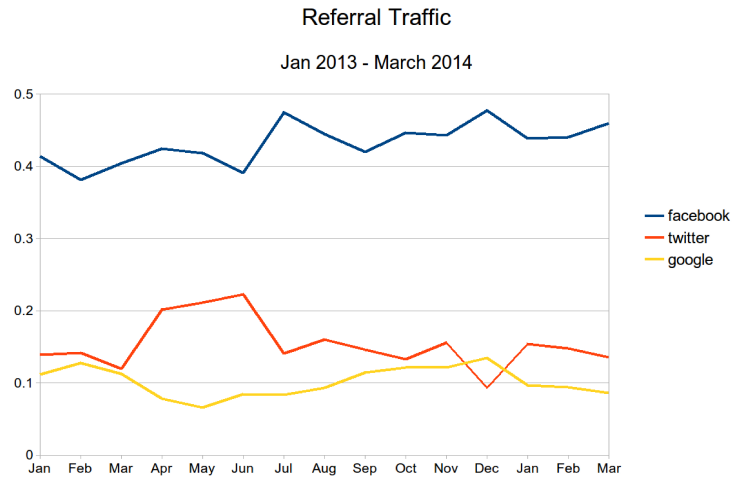

Overall, about 3.5 per cent of all the visits to the website start at the homepage displaying the trending petitions information. If we look at the traffic sources of this 3.5 per cent, we observe a very different pattern compared to the overall visits; only 6 per cent of this 3.5 per cent is sourced from Facebook (equally from mobile and desktop) while Twitter is only responsible for 2 per cent of this direct traffic to the front page. About 44 per cent of the visits starting at the homepage originated from Google and 30 per cent of them were direct visits (by users who had typed the web address of the petition site directly into their browsers, bookmarked the page, or clicked a link in an email). That many users starting at the homepage come from Google is not surprising, considering that the first search result in Google for the keyword 'petition' directs to this page (at least for users in the UK).

These observations contribute to a picture of aimless petitioners: they either just go to the website directly or look for the petition website in Google, without aiming at a certain petition. Apart from this group who visit the homepage and trending petition information directly, almost 10 per cent of all the first clicks within the website (that is, by users who arrived at a page other than the homepage on the site) lead users to the homepage (with the trending petition information). This is followed by 7%, 8%, and 9% of $2^{nd}$, $3^{rd}$, and $4^{th}$ interactions of the users with the website. Eventually, a consistent number of site visits at each stage directs to the trending petition information on the homepage. This also includes the users who sign a petition and then visit the homepage. Overall, this led to 2.35 million visits to the homepage (1.40 million unique visitors) out of the 63.6 million page views on the site overall (47.1 million unique page visits) during the period of traffic analysis. There were about 5.9 million visit sessions with at least one interaction with the site content (there were 21.8 million sessions in total, out of which 15.9 million ended without the user clicking on anything) and about 20 per cent of all these sessions passed through the homepage at least once during the first 10 clicks.

Using these numbers it is possible to estimate the theoretical maximum amount of change in the Gini coefficient of the signatures distribution that we would expect to see due to the social influence. We denote this change $\Delta g$, and note that $\Delta g$ cannot be larger than $(1 - g_0)\frac{\Delta N}{N}$ where $g_0$ is the Gini coefficient without social information and $\frac{\Delta N}{N}$ is the fraction of influenced visitors. By plugging in the values of $g_0 = 0.82$ and $\frac{\Delta N}{N} = 0.20$ (one fifth of total visits, which we compute by assuming all users to the homepage are influenced by the trending information and sign one of the top few petitions), we find that the theoretical value for $\Delta g$ is 0.04. The amount of change in the Gini Coefficient that we observed in the signature data is 0.03, indicating that the effects of the local treatment is very strong, although due to the relatively small size of the treatment group (visitors who see the homepage) the overall effect on petitions as a whole is not extremely large (about 3%). The maximum value for $\Delta g$ is limited not only by the relatively small size of the treatment group ($\frac{\Delta N}{N}$), but also by the large starting value of the Gini coefficient even before the introduction of the trending petition information (reflected in $1 - g_0$ factor). Since the Gini coefficient was already large even before the trending information was introduced ($g_0 = 0.82$), it is rather unlikely to increase a large amount. If $g_0$ had been smaller before the change in the design (and thus $1 - g_0$ would have been larger), then the effect of the design change could have been even stronger.

**Conclusions**

This paper has analysed some new data on political behaviour to investigate social information effects in digital environments, suggesting that some citizens are using the internet for generalized (rather than issue-based) civic engagement and showing how the design of internet-based participation platforms can have significant effects on individuals' political behaviour. We have examined the results from a natural experiment, where the effect of platform design changes can be observed and explained. We have verified the importance of social information, by showing that information about the participation of others—trending information—can have an important effect on those who are exposed to it. Although the effect size for the complete population of petition signers was small, we identified a group of 'aimless petitioners', who deliberately sought out the homepage of the petitions platform where the information was displayed, knowing that they wanted to sign a petition but uncertain as to the issue on which they wanted to petition. For this group, the effect of the information was strong. Now that so much of political participation takes place in digital contexts, where people are invited to make 'micro-donations' of time and effort to political causes as they go about their daily lives on social media, social information is more abundant and we can expect that social information effects of this kind will be more prevalent than in offline contexts, where decisions to participate tend to be larger, more 'lumpy' and harder to influence (and therefore less likely to be influenced by weak informational cues).

Our findings also suggest that social information in the form of trending information can exacerbate turbulence in political mobilization. The most popular petitions receive more signatures than one would expect just by virtue of their popularity, and in this zero-sum race for collective attention, those for which there is no evidence of popularity receive fewer signatures than their popularity would indicate, even though their actual signature levels may be very close to those that appear in the trending box. In this way we have observed the same kind of effects that Salganik et al. (2006, 2009) have observed for cultural markets, where experimental subjects were shown varying information about the popularity of songs and were asked to rank them. Subjects were more likely to rate highly those songs for which there was evidence of popularity, meaning that the information injected a source of instability into cultural markets. We can expect that this finding could be generalized to other forms of online collective action, as these kinds of popularity indicators are present in some form by default on most social media. On Twitter for example, trending 'hashtags' (topics currently mentioned in many tweets) are shown on the main screen of the user. Moreover, a recent change to Twitter has introduced a 'Top Tweet' facility, which means that the most retweeted tweets produced by those users that a user follows are shown at the top of their Twitter feed. On Facebook, any post will show information about the number of 'likes' it has received, YouTube videos show the number of views, and so on.

All this information about popularity is likely to have the same kind of reinforcement effect on any collective action that is disseminated via social media, as is the case for most campaigns and mobilizations. For example, the analytics data we used at the end of the paper shows that around two thirds of all visits to the petitions site arrived via either Twitter or Facebook, and the burgeoning array of additional social media sites such as Pinterest, Tumblr and Instagram are likely to increase the importance of social media for petitions websites.

Further research could investigate the influence of the design of these alternative platforms, and how social information is presented within that design, on the likelihood of petitions being disseminated and, ultimately, signed. We might also investigate the effects of trending information (that is, rate of change in popularity) relative to the overall popularity of a petition, something that we are already exploring with the dataset analysed here. However, given that both pieces of information, number of signatures per hour and total signatures so far, are provided on the homepage (albeit in different font sizes), it may be difficult to disentangle their relative effects in these data.

The implications of the growing influence of social information are twofold. The first is that we will see more social information generated participation, which contrasts with the decline of in mobilization in other contexts, such as through political parties and partisan alignment and other political institutions. This finding—if replicated across time and place—may have implications for political systems and their legitimacy. The second impact is that it is probable that we will see mobilization of a different kind of citizen from before. We know that social information is likely to appeal to those with certain types of personality—agreeable people, for example—while others (such as extraverts) are much more resistant to its influence (Margetts et al., 2013), which would indicate that the individuals mobilized might be statistically different from the non-mobilized. However, it is likely that those already mobilized will have been brought into politics by other factors than social information so we expect the newly mobilized to be different from the existing mobilized, for example having less interest in politics and being less socially exclusive than those already involved (John, 2009, although see also Escher, 2011).

Notwithstanding the distinctive characteristics of people who are encouraged to participate by social information, these findings might generate some cautious optimism with respect to political engagement. The finding that some people are keen to participate politically in a generalized (rather than an issue specific) way might be taken as encouraging at a time when so much political commentary bemoans both the time that certain groups (particularly young people) spend online on social media, and growing political disengagement amongst these groups. We have some modest evidence here that in fact, significant numbers of people are looking for political causes as they go about their lives online. In combination with other evidence that making small internet-based contributions to political causes seems to lead on to more substantive contributions (for example in the Obama election campaigns, see Wilcox, 2008), findings like this speak against those who foresee an unstoppable decline in political participation and engagement.

**Acknowledgement**
This work was supported by the Economic and Social Research Council [grant number RES-051-27-0331].

**Bibliography**
Andreoni, J. (2006) 'Philanthropy' in L.A. Gerar-Varet, S-C. Kolm and J.Mercier Ythier (eds.) *The Handbook of Giving, Reciprocity and Altruism*, Handbooks in Economics, Amsterdam: North-Holland.
Andreoni J. (2006) 'Leadership giving in charitable fund-raising', *Journal of Public Economic Theory* 8(1):1–22, DOI 10.1111/j.1467-9779.2006.00250.x
Borge, R. and Cardenal, A. (2011) 'Surfing the Net: A pathway to participation for the politically uninterested?', *Policy and Internet*, 3 (1).
Boyd, D. (2009) 'Streams of Content, Limited Attention: The Flow of Information Through Social Media', Web2.0 Expo, New York, 17th November.
Buscher, G., Cutrell, E., and Ringel Morris, M. (2009). What do you see when you're surfing?: Using eye tracking to predict salient regions of web pages. In Proceedings of the SIGCHI Conference on Human Factors in Computing Systems (CHI '09). ACM, New York, NY,

USA, 21-30. DOI=10.1145/1518701.1518705 http://doi.acm.org/10.1145/1518701.1518705

Carpenter, D. (2003) "http://people.hmdc.harvard.edu/~dcarpent/petition-recruit-20040112.pdf"

Chadwick, A. 2012. "Web 2.0: New challenges for the study of e-democracy in an era of informational exuberance". In *Connecting democracy: Online consultation and the flow of political communication*, Edited by: Coleman, S. and Shane, P. M. 45–75. Cambridge, MA: MIT Press.

Cialdini R. and Goldstein, J. (2004) 'Social influence: Compliance and conformity', *Annual Review of Psychology* 55:592–621

Croson R.T., Shang, J. (2008) The impact of downward social information on contribution decisions. *Experimental Economics* 11(3): 221–233

Dunning, T. (2012) *Natural Experiments in the Social Sciences: a Design-based Approach*. Cambridge University Press.

Escher, T. 2011. *Writetothem.com: Analysis of users and usage for UK Citizens Online Democracy*, London, , England: My Society.

Fischbacher, U. and Gächter, S. (2010) 'Social Preferences, Beliefs, and the Dynamics of Free Riding in Public Goods Experiments', *The American Economic Review,* 100(1): 541–556.

Fox, R. (2012) What Next for E-petitions. London: Hansard Society.

Frey, B.S. and Meier, S. (2004) 'Social Comparisons and Pro-social Behavior: Testing "Conditional Cooperation" in a Field Experiment', *American Economic Review.* 94: 1717–1722.

Gelertner, D. (2013) 'The End of the Web, Search and Computer as we know it', *Wired*, 1st February.

Gerber, A., Green, D. Larimer, C. (2008) 'Social pressure and voter turnout: evidence from a large-scale field experiment', *American Political Science Review* 102(1): 33–48.

Goldstein, N., Cialdini, R. and Griskevicius, V. (2008). 'A room with a viewpoint: Using social norms to motivate environmental conservation in hotels', *Journal of consumer Research* 35.3: 472-482.

Hale, Scott A., Helen Margetts, and Taha Yasseri (2013) "Petition growth and success rates on the UK No. 10 Downing Street website." *Proceedings of the 5th Annual ACM Web Science Conference*. ACM.

John, P. (2009), 'Can Citizen Governance Redress the Representative Bias of Political Participation?, *Public Administration Review*, 69: 494-503.

Jungherr, A. and Jurgens, P. (2012) 'The Political Click: Political Participation through E-Petitions in Germany, *Policy and Internet*, 2(4) pp. 131-165,

Lindner, R. and Riehm, U. (2009) Electronic petitions and institutional modernization: International parliamentary e-petition systems in comparative perspective. *JeDEM—eJournal of eDemocracy and Open Government*, 1: 1–11.

Lupia, Arthur, and Gisela Sin (2003) Which public goods are endangered?: How evolving communication technologies affect the logic of collective action." *Public Choice* 117.3-4: 315-331.

Margetts, H., John, P., Escher, T., and Reissfelder, S. (2011) 'Social Information and Political Participation on the Internet: an Experiment', *European Political Science Review,* 3(3): 321:344.

Margetts, H. Z., John, P., Hale, S. A. and Reissfelder, S. (2013) Leadership without Leaders? Starters and Followers in Online Collective Action. Political Studies. doi: 10.1111/1467-9248.12075

Margetts, H.Z., John, P., Hale, S.A. and Yasseri, T. (2015) *Chaotic Pluralism: Social Media and Collective Action*., forthcoming.

Marwell, Gerald, and Pamela Oliver (1993)*The critical mass in collective action*. Cambridge University Press.

Narayanan, Sridhar, and Kirthi Kalyanam. *Measuring position effects in search advertising: A*

*regression discontinuity approach*. Working Paper, 2011.

Pfeffer J. and Salancik G.R. (1978) *The External Control of Organizations: a Resource Dependence Perspective*. New York: Harper and Row.

Salancik, G. and Pfeffer, J., (1977) 'An examination of need-satisfaction models of job attitudes', *Administrative Science Quarterly* **22:** 427–456.

Salganik, M., Dodds, P. and Watts, D. (2006), 'Experimental study of inequality and unpredictability in an artificial cultural market', *Science* 311(5762): 854–6.

Salganik, M. and Watts, D. (2009) 'Web-based experiments for the study of collective social dynamics in cultural markets', *Topics in Cognitive Science* 1: 439-468.

Schelling, T. (2006) Micromotives and Macrobehaviour. New York. WW Norton. 1st edition 1978.

Schultz, P. W. (1999) 'Changing behavior with normative feedback interventions: a field experiment of curbside recycling', *Basic and Applied Social Psychology* 21: 25-36.

Shang, J. and Croson, R. (2009) 'A field experiment in charitable contribution: the impact of social information on the voluntary provision of public goods', *The Economic Journal* 119(540): 1422-1439.

Smith, G. (2009) *Democratic innovations: Designing institutions for citizen participation*, Cambridge, England: Cambridge University Press.

Whyte, A., Renton, A., and Macintosh, A. (2005). *e-Petitioning in Kingston and Bristol Evaluation of e-Petitioning in the Local e-Democracy National Project*. International Teledemocracy Centre, Napier University.

Wilcox, C. (2008) 'Internet Fundraising in 2008: a New Model', The Forum, 6(1) and other articles in this issue.

Wright, S. (2012) 'Assessing (e-)Democratic Innovations: "Democratic Goods" and Downing Street E-Petitions', Journal of Information Technology and Politics 9(4).

Yasseri, Taha, Scott A. Hale, and Helen Margetts (2014) "Modeling the Rise in Internet-based Petitions." *arXiv preprint arXiv:1308.0239*,